# Abnormality Detection in Musculoskeletal Radiographs with Convolutional Neural Networks(Ensembles) and Performance Optimization

Dennis BANGA[1] Peter WAIGANJO[1]
School of Computing & Informatics, University of Nairobi, Box 30197, Nairobi, Kenya
Email: dennisbanga@gmail.com, waiganjo@uonbi.ac.ke

**Abstract**: Musculoskeletal conditions affect more than 1.7 billion people worldwide based on a study by Global Burden Disease, and they are the second greatest cause of disability[1,2]. The diagnosis of these conditions vary but mostly physical exams carried out and image tests. There are few imaging and diagnostic experts while there is a huge workload of radiograph examinations which might affect diagnostic accuracy.

We built machine learning models to perform abnormality detection using the available musculoskeletal public dataset [3]. Convolutional Neural Networks (CNN) were used as are the most successful models in performing various tasks such as classification and object detection [4]. The development of the models involved theoretical study, iterative prototyping, and empirical evaluation of the results.

The current model, 169 layer DenseNet, by Pranav et al.(2018) on the abnormality detection task, the performance was lower than the worst radiologist in 5 out of the 7 studies, and the overall model performance was lower than the best radiologist.

We developed the *ensemble200 model* which scored 0.66 Cohen Kappa which was lower than the DenseNet model (Pranav et al, 2018) but the model performance with the F1 score outperforms the DenseNet model and its Cohen Kappa score variability with the different studies is lower as the best cohen kappa score on the upper extremity studies is 0.7408 (Wrist) and the lowest is (0.5844) hand. The *ensemble200 model* outperformed DenseNet model on the finger studies with a Cohen Kappa score of 0.653 showing reduced performance variability on the model performance.

## 1. Introduction

According to WHO, musculoskeletal conditions are categorized in more than 150 diagnosis affecting the musculoskeletal (locomotor) system[5]. This includes muscles, bones tissues (tendons and ligaments) and joints. Some musculoskeletal conditions such as fractures are short lived while others are lifelong with the patients having prevalent pain or permanent disability.

In Kenya, a total of 158,055 musculoskeletal cases were reported in the KHIS (Kenya Health Information System- Ministry of Health) in 2017. In addition, the total number of experts in diagnostics and imaging is 376 according to iHRIS (integrated Human Resource Information System for the Ministry of Health), and the total radiology examinations reported in KHIS were 869,577 in 2017 [6,7]. This clearly shows a shortage of experts based on the level of workload and might affect the diagnostic accuracy.

Stanford University- Department of Computer Science, Medicine, and Radiology introduced a public dataset MURA of musculoskeletal radiographs from Stanford Hospital which is the largest dataset with 40,561 images from 14,863 upper extremity studies [3]. The DensNet model developed to detect the abnormalities has a performance that is lower than the worst radiologist in 5 out of the 7 studies, and the model's overall performance was not comparable to the best radiologist performance. The DenseNet model achieved an AUROC of 0.929 (0.815 sensitivity and 0.887 specificity).

To improve on the performance consistency and generalization of the convolutional neural network model, an ensemble model was built with the trained single models(DenseNet201, MobileNet, NASNetMobile) on the Musculoskeletal radiographs which involved evaluating different ensemble combination performance and selecting the best using the copen kappa statistic.

2. Related Work

2.1 Pneumonia Detection on Chest X-Rays with Deep Learning (ChexNet) [8]

ChexNet is a 121 layer convolutional neural network developed by Rajpurkar P. et al. (2017) for detecting pneumonia from chest X-rays. The model was trained on over 100,000 Chest X-ray frontal view dataset and containing 14 other diseases. The model detected pneumonia from the X-ray images at a performance exceeding radiologist.
The DenseNet Convolutional Neural Network architecture was utilized as it improves gradient and information flow in the network. The fully connected layer has a single output,and then sigmoid function is applied. The weights of the network are initialized with weights of a pre-trained model on ImageNet dataset and then the network is trained end to end.

2.2 Abnormality Detection in Musculoskeletal Radiographs Dataset [3]

A model to detect and localize abnormalities in musculoskeletal radiographs developed by Rajpurkar P. et al. (2017). It is a 169-layer DenseNet Convolutional Neural Network model that was trained on a musculoskeletal radiographs dataset containing 40,561 images from 14,863 studies. The model performance was an Area Under Receiver Operating characteristic(AUROC) of 0.815, sensitivity and specificity of 0.887.
The Rajpurkar P. et al. (2017) DenseNet Model performance was comparable to the best radiologist performance on the wrist and finger studies, but the performance is lower than best radiologist performance in detecting abnormalities on elbow, forearm, hand, humerus, and shoulder studies. Hence, a gap exists in terms of model performance utilizing the DenseNet model. As per the review of the different convolutional neural network architectures and deep learning methods, performance can be improved in a number of ways by applying some techniques.

### 2.3 Hemorrhage detection in CT Scans with Deep learning (RADNET) [9]

Recurrent Attention DenseNet Model detects hemorrhage in computed tomography(CT) scans with an accuracy of 81.82% comparable to radiologists. The model utilizes DenseNet Architecture 40 layers along with components of slice level predictions and recurrent neural network layer. The model is trained on 185 brain CT scans, and 67 scans for validation and 77 for testing. The model utilizing the DenseNet architecture, and its performance is comparable to radiologist.

### 2.4 Dermatologist-level classification of skin cancer with deep neural networks [10]

A convolutional neural network model for classifying skin lesions developed by Esteva A. et al. (2017). The model was trained on a dataset composed of 129,450 clinical images with 2,032 different diseases. The network was trained end to end, and the model performance was comparable to expert performance.

### 2.5 Abnormality Detection in Mammography using Deep Convolutional Neural Networks [11]

A Convolutional neural network model for classifying and localizing masses on mammogram images. A ResNet CNN architecture was used for calculating class activation maps. The model utilized transfer learning and fine tuned on pre-trained CNNs of image patches that are cropped. A full mammography image is fed as the input of the CNN tuned on patch images and class activation maps computed for localizing abnormalities.

## 3. Objectives

The main objective of the study was to develop a convolution neural network model that automatically detects abnormalities and normalities in musculoskeletal radiographs with improved model performance.

The specific objectives of the study were:
- To preprocess the musculoskeletal dataset for training on the different CNN architectures.
- To design and evaluate an ensembled convolutional neural network architecture for training the musculoskeletal radiographs and perform performance tuning.
- To evaluate the generalization of the model with the test dataset.
- To perform abnormality localization on the cases automatically detected for model interpretability.

- To design and develop a web-based interface for abnormality detection with the evaluated models.

## 4. Methodology

In the study, different available convolutional neural network architectures were explored and reviewed based on their document strengths as top-1 accuracy, top-5 accuracy, the size of the model, the number of parameters and the depth of the network and this was documented. The identified CNN architectures were used in building the single models and ensembles. The MURA dataset was preprocessed into a structure and format that would allow training and validation on the different CNN models.

Through iterative prototyping, different single models and their ensembles were designed, implemented, trained and evaluated using empirical metrics (Precision, Recall, Sensitivity, Specificity, AUROC, Accuracy and Copen Kappa Static). The training of the models was fine-tuned by adjusting the hyper-parameters and the model structure. The empirical evaluation results of the different models were documented and contrasted with the radiologist performance in the MURA dataset.

In addition, model interpretability and localization of the model predictions was performed using gradient class activation maps.

An interactive web interface was developed to demonstrate the utilization of the evaluated models in a clinical setup.

### 4.1 Source of Data

The musculoskeletal public dataset for this research is made up of 14,863 upper extremity studies from 12,173 patients, and a total of 40,561 multi-view radiographic images from stanford hospital [3]. The dataset consists of study types of the finger, elbow, hand, humerus, forearm shoulder and wrist. The studies are labelled abnormal or normal by radiologists manually, and split into training and validation sets for evaluation.

| Study | Train Normal | Train Abnormal | Validation Normal | Validation Abnormal | Total |
|---|---|---|---|---|---|
| Elbow | 1094 | 660 | 92 | 66 | 1912 |
| Finger | 1280 | 655 | 92 | 83 | 2110 |
| Hand | 1497 | 521 | 101 | 66 | 2185 |
| Humerus | 321 | 271 | 68 | 67 | 727 |
| Forearm | 590 | 287 | 69 | 64 | 1010 |
| Shoulder | 1364 | 1457 | 99 | 95 | 3015 |
| Wrist | 2134 | 1326 | 140 | 97 | 3697 |
| Total No. of Studies | 8280 | 5177 | 661 | 538 | 14656 |

Table 1. Musculoskeletal radiography dataset of 14,863 studies [3].

## 4.2 Data Collection

The musculoskeletal public dataset of 40,561 radiographic image studies collected from Picture Archive and Communication Systems of Stanford Hospital, and the images are HIPAA-compliant. Each of the studies were manually labelled as normal or abnormal by radiologists who are boardcerified. The dataset was split into training and validation sets, with no overlap of the datasets.

To understand the clinical workflow of patients, and the activities carried out in a radiology department a visit was made in two local hospital in Kenya (St Mary's Hospital Langata & Mariakani Cottage Hospital South B) . Through an informal interview with the radiologist technicians and observing the clinical workflow, in the two facilities. The process is digitized and the patient records are stored in an electronic medical record system.

## 4.3 Conceptual Model

The input data to the model is the labelled normalized upper extremity radiograph image views. The model is made up of three selected convolutional neural network models trained on the MURA dataset then stacked to an ensemble model. The output of the ensemble model is the averaged scores of the prediction, and the probability of abnormality. Localization of the abnormalities utilized the gradient class activation maps. Network training was done end to end,and the model weights initialized with pre-trained imagenet weights. Fine tuning the network hyper parameters was done continuously to optimize the model performance.

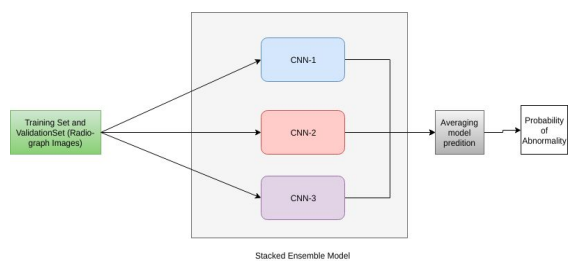

Figure 1. Ensembled Model

## 4.4 Data Analysis

The MURA dataset was utilized for the evaluation of the models. The training and the validation dataset was utilized and the quantitative performance metrics measured were the accuracy, specificity, sensitivity, precision, recall, F1-Score, AUROC (Area Under Receiver Operating Characteristics), and Cohen's Kappa statistic to compare model and radiologist score as mapped out by Rajpurkar P. et al. (2017). Model interpretation was done using Gradient-weighted Class Activation mapping that produces visual explanations of the prediction results.

## 4.5 Model Combination

In the design of ensemble models from single models, different combinations were designed and evaluated. For the ensembles, an additional averaging layer was added to provide the average prediction probability.

Tensorflow playground tool was used to simulate different neural network sizes on a binary classification problem and adjusting the hyper parameters.

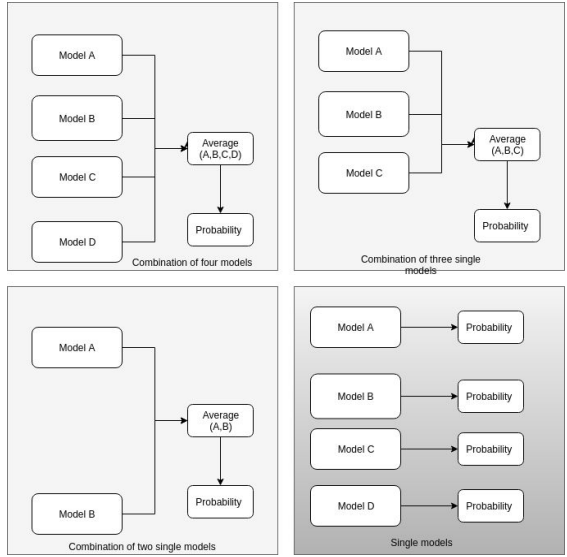

Figure 2. Ensemble Combinations and Single Model.

## 4.6 Network Training

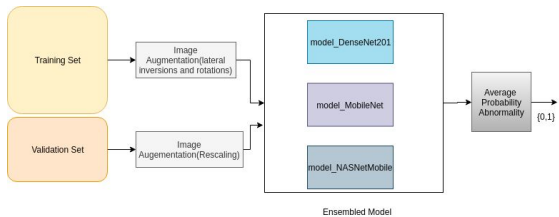

Figure 3. Ensembled Architectural Design.

Four different single models were trained on the MURA dataset, namely: DenseNet201, Xception, MobileNet, NASNetMobile models. The models were initialized with pretrained imagenet weights, then an additional layer of global average pooling was added to the network that gets an input of a 4 dimensional tensor and outputs a 2 dimensional tensor. Then a sigmoid function was applied on the last layer.

$$\sigma(z) = \frac{1}{1 + e^{-z}}$$

Figure 4. Sigmoid function Equation. It converts the input value to a value between 0 and 1.

On the training dataset, image normalization and transformations were applied to the images. The training and the validation set images were normalized by rescaling with a factor of (1/255), this is to reduce the image target values to be between 0 and 1 as not to have too high values to process. For the training dataset, the images were transformed by performing random lateral inversions and rotations. (horizontal flips, rotation_range=45). Both the training and validation set images were scaled to variable-sized images of 224*224. This image augmentation operations were important to avoid overfitting and provide better generalization of the models.

The class balancing technique were applied to the classes to handle the imbalanced classes,as the number of normal musculoskeletal radiographic studies were 9,045 while the number of abnormal studies were 5,818. A balanced heuristic was used to weight the two classes.

The network for the four models were independently trained end to end using the binary cross entropy loss function and the Adam optimizer.

$$H_p(q) = -\frac{1}{N} \sum_{i=1}^{N} y_i \cdot log(p(y_i)) + (1 - y_i) \cdot log(1 - p(y_i))$$

Figure 6. Binary cross entropy loss function.

The label y is 1 for normal radiographic images and 0 for abnormal radiographic images, while p(y) is the predicted probability for the normal radiographic image.

The Adam optimizer is a 1st order gradient optimizer for stochastic objective functions, and was used as it is computationally efficient, has lower memory requirement, invariant to diagonal rescaling, and favourable for both problems with large data and parameters[18].

In the training the models, the Adam optimizer parameters applied were as follows (learning rate=1-e4, beta_1=0.9, beta_2=0.999, epsilon=None, decay=1-e4, amsgrad=False)

The network was trained in mini batches of 32. After the models were trained, and the validation evaluation results were satisfactory, the models were combined into different designed ensembles.

## 4.7 Ensemble Combinations

Different ensemble combinations were assembled and evaluated. The ensembles were configured by adding an averaging layer of the sigmoid output layer of the single models.

![Figure 7 code](ensemble code)

```python
def ensemble(models, model_input, modelName):
    outputs = [model.outputs[0] for model in models]
    y = Average(name='average_predictions')(outputs)
    model = Model(model_input, y, name=modelName)
    return model
```

**Ensemble Variants**

```python
modelensemble = [model_DenseNet201, model_MobileNet, model_NASNetMobile]
modelA=[model_NASNetMobile,model_Xception,model_MobileNet]
modelB=[model_NASNetMobile, model_MobileNet]
modelC=[model_Xception,model_MobileNet]
modelD=[model_Xception,model_NASNetMobile]
modelE=[model_Xception,model_DenseNet201]
modelF=[model_Xception,model_DenseNet201, model_MobileNet]
modelAll=[model_DenseNet201,model_NASNetMobile,model_Xception,model_MobileNet]

model_Ensemble = ensemble(modelensemble, model_input, "ensemble")
model_EnsembleA = ensemble(modelA, model_input, "ensembleA")
model_EnsembleB = ensemble(modelB, model_input, "ensembleB")
model_EnsembleC = ensemble(modelC, model_input, "ensembleC")
model_EnsembleD = ensemble(modelD, model_input, "ensembleD")
model_EnsembleE = ensemble(modelE, model_input, "ensembleE")
model_EnsembleF = ensemble(modelF, model_input, "ensembleF")
model_EnsembleAll = ensemble(modelAll, model_input, "ensembleAll")
```

Figure 7. Ensemble Function and the Ensemble Varying Combinations

```
avg_pool_densenet201 (GlobalAve  (None, 1920)    0      relu[0][0]
avg_pool_mobilenet (GlobalAvera  (None, 1024)    0      conv_pw_13_relu[0][0]
avg_pool_nasnetmobile (GlobalAv  (None, 1056)    0      activation_188[0][0]
predictions_densenet201 (Dense)  (None, 1)       1921   avg_pool_densenet201[0][0]
predictions_mobilenet (Dense)    (None, 1)       1025   avg_pool_mobilenet[0][0]
predictions_nasnetmobile (Dense  (None, 1)       1057   avg_pool_nasnetmobile[0][0]
average_predictions (Average)    (None, 1)       0      predictions_densenet201[0][0]
                                                        predictions_mobilenet[0][0]
                                                        predictions_nasnetmobile[0][0]
========================================================================================
Total params: 25,824,567
Trainable params: 25,536,885
Non-trainable params: 287,682
```

Figure 8. Ensemble Model Implementation. (Average predictions layer)

## 5. Results and Analysis

The evaluations were based on the following metrics Cohen Kappa, F1 Score, accuracy, precision, recall, sensitivity, ROC Score and specificity, compared with MURA paper v2 and v4 result published.[3]

Based on the evaluation of the different combinations of ensembles (DenseNet201, mobileNet, NASNETmobile and Xception) and the single models. The evaluation results were as the summary below.

| Overall Models Evaluation - Per Encounter Metrics | | | | | |
|---|---|---|---|---|---|
| **Models** | *Accuracy* | *F1 Score* | *Precision* | *Recall/Sensitivity* | Rank |

| Model | | | | | |
|---|---|---|---|---|---|
| (Single)MobileNet | 0.67 | 0.71 | 0.7 | 0.73 | 3 |
| (Single)NASMobile | 0.43 | 0.52 | 0.48 | 0.57 | |
| (Single)Xception | 0.53 | 0.58 | 0.59 | 0.59 | |
| (Single)DenseNet201 | 0.29 | 0.37 | 0.36 | 0.38 | |
| EnsembleA [NASN, Xcep,MobileN] | 0.57 | 0.63 | 0.6 | 0.67 | 5 |
| EnsembleB [NASN, MobileN] | 0.27 | 0.37 | 0.35 | 0.39 | |
| EnsembleC [Xcep,MobileN] | 0.37 | 0.44 | 0.43 | 0.44 | |
| EnsembleD[Xcep, MobileN] | 0.65 | 0.7 | 0.67 | 0.73 | 4 |
| EnsembleE [Xcep,Dense] | 0.71 | 0.75 | 0.72 | 0.77 | 2 |
| EnsembleF [Xcep, Dense, MobileN] | 0.21 | 0.3 | 0.3 | 0.31 | |
| EnsembleALL | 0.44 | 0.52 | 0.49 | 0.55 | |
| Ensemble200[Dense,MobileN, NASN] | 0.83 | 0.86 | 0.81 | 0.92 | 1 |

Table 3. Evaluation Results for the Single models and Ensembles

**Overall Models Evaluation - Per Encounter Metrics**

| Models | *Specificity* | *ROC_Score* | *Cohen Kappa* | Evaluation Acc. | Rank |
|---|---|---|---|---|---|
| (Single)MobileNet | 0.61 | 0.67 | 0.34 | 0.7726 | 3 |
| (Single)NASMobile | 0.25 | 0.41 | -0.18 | 0.7676 | |
| (Single)Xception | 0.47 | 0.53 | 0.06 | 0.7917 | |
| (Single)DenseNet201 | 0.18 | 0.28 | -0.44 | 0.7591 | |
| EnsembleA [NASN, Xcep,MobileN] | 0.46 | 0.56 | 0.13 | 0.8061 | 5 |
| EnsembleB [NASN, MobileN] | 0.13 | 0.26 | -0.49 | 0.7932 | |
| EnsembleC [Xcep,MobileN] | 0.29 | 0.36 | -0.27 | 0.8033 | |

| Model | | | | | |
|---|---|---|---|---|---|
| EnsembleD[Xcep, MobileN] | 0.56 | 0.64 | 0.29 | 0.8051 | 4 |
| EnsembleE [Xcep,Dense] | 0.63 | 0.7 | 0.41 | 0.7998 | 2 |
| EnsembleF [Xcep, Dense, MobileN] | 0.1 | 0.2 | -0.6 | 0.8008 | |
| EnsembleALL | 0.3 | 0.43 | -0.15 | 0.8076 | |
| Ensemble200[Dense,MobileN, NASN] | 0.73 | 0.82 | 0.66 | 0.7973 | 1 |

Table 4. Evaluation Results for the Single models and Ensembles

From the results, the best performing model was ensemble200[DenseNet201, MobileNet, NASNETMobile] that had a Kappa Score of 0.66, precision of 0.81, recall/sensitivity of 0.92, ROC score of 0.82, and F1 score of 0.86. The second performing model was ensembleE[Xception, and DenseNet201] with Kappa of 0.41, the third was a single model (MobileNet) with kappa of 0.34. Of the 8 ensembles 4[Ensemble200, EnsembleE, EnsembleD, EnsembleA] had an accuracy of higher than 0.55 while the only we had one single model.

We then compared our model performance with the MURA(v4) (Pranav R et al, 2018) results. The evaluation metric was the Cohen Kappa(measures the level of agreement between two evaluators), precision and recall. The overall performance of our model was Kappa 0.66 (Precision-0.81, Recall 0.92), which was lower compared to the MURA(v4) model with a score of 0.705 (0.700, 0.710) and the best radiologist with a score of 0.778 (0.774, 0.782). On the upper extremity studies of the Finger, our model outperformed all the radiologist and the MURA(v4) model.

For our model, the best score was on the wrist studies 0.7408 (0.8581, 0.9500, which was also the best scoring study for the radiologist 2 (0.931 (0.922, 0.940)), radiologist 3 (0.931 (0.922, 0.940)), and the MURA(v4) model (0.931 (0.922, 0.940)).

For the worst performing study in our model was the Hand with a score of (0.5844 (0.7778, 0.9703)) while for the MURA(v4) model and the radiologist was the finger studies. The best finger scores for this was 0.410 (0.358, 0.463) comparing to our model score of 0.653 (0.7870, 0.9239) on the finger studies. The table below shows the detailed comparisons.

| MURA(v4) Paper - Metric Cohen Kappa, (Precision, Recall) - Per Encounter Metrics | | | | | |
|---|---|---|---|---|---|
| **Upper Extremity Studies** | Radiologist 1 | Radiologist 2 | Radiologist 3 | Paper Model | **Our Model** |
| Elbow | 0.850 (0.830, 0.871) | 0.710 (0.674, 0.745) | 0.719 (0.685, 0.752) | 0.710 (0.674, 0.745) | *0.617 (0.8182, 0.8804)* |
| Finger | 0.304 (0.249, 0.358) | 0.403 (0.339, 0.467) | 0.410 (0.358, 0.463) | 0.389 (0.332, 0.446) | *0.653 (0.7870, 0.9239)* |
| Forearm | 0.796 (0.772, 0.821) | 0.802 (0.779, 0.825) | 0.798 (0.774, 0.822) | 0.737 (0.707, 0.766) | *0.6954 (0.7753, 1.0000)* |
| Hand | 0.661 (0.623, 0.698) | 0.927 (0.917, 0.937) | 0.789 (0.762, 0.815) | 0.851 (0.830, 0.871) | *0.5844 (0.7778, 0.9703)* |
| Humerus | 0.867 (0.850, 0.883) | 0.733 (0.703, 0.764) | 0.933 (0.925, 0.942) | 0.600 (0.558, 0.642) | *0.5995 (0.7595, 0.8824)* |
| Shoulder | 0.864 (0.847, 0.881) | 0.791 (0.765, 0.816) | 0.864 (0.847, 0.881) | 0.729 (0.697, 0.760) | *0.6597 (0.8367, 0.8283)* |
| Wrist | 0.791 (0.766, 0.817) | 0.931 (0.922, 0.940) | 0.931 (0.922, 0.940) | 0.931 (0.922, 0.940) | *0.7408 (0.8581, 0.9500)* |
| Overall | **0.731 (0.726, 0.735)** | **0.763 (0.759, 0.767)** | **0.778 (0.774, 0.782)** | **0.705 (0.700, 0.710)** | ***0.66 (0.81, 0.92)*** |

Table 5. Evaluation Result Comparison using the Cohen Kappa Metric with our best Ensemble200 model (This evaluation results were based on the validation set data.)

5.1 Model Interpretation

The Grad Class Activation Maps were applied to perform model interpretation of the prediction results of the models.

Below are the results of model interpretation using Grad CAM.

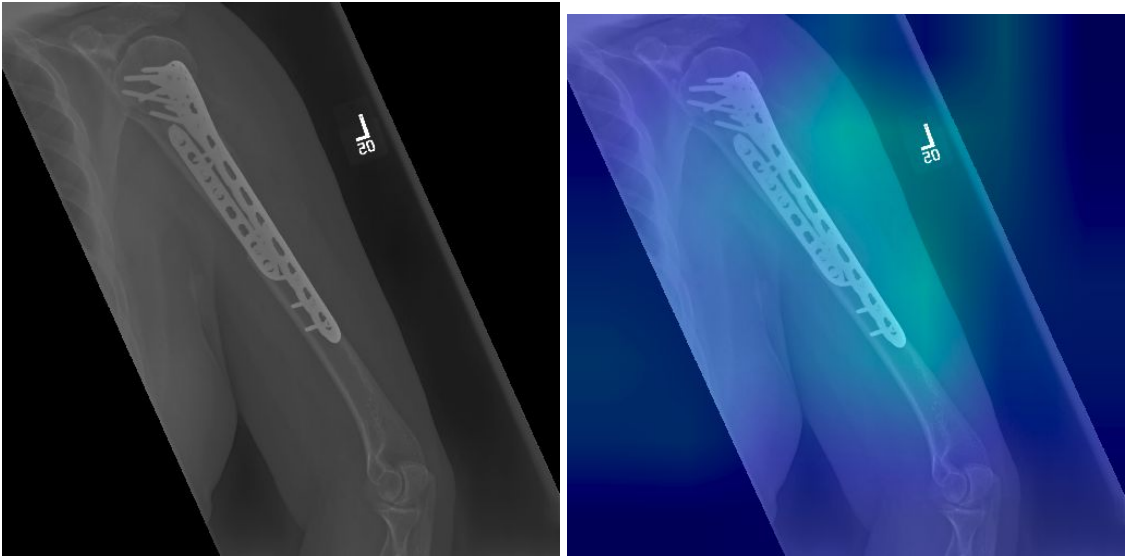

Figure 10. Abnormal Humerus Predicted using Ensemble200 as abnormal(Positive, 0) with a score of (0.05). Image at the top is the input to the model, and the image at the bottom is the output of the Grad CAM algorithm with highlighted regions.

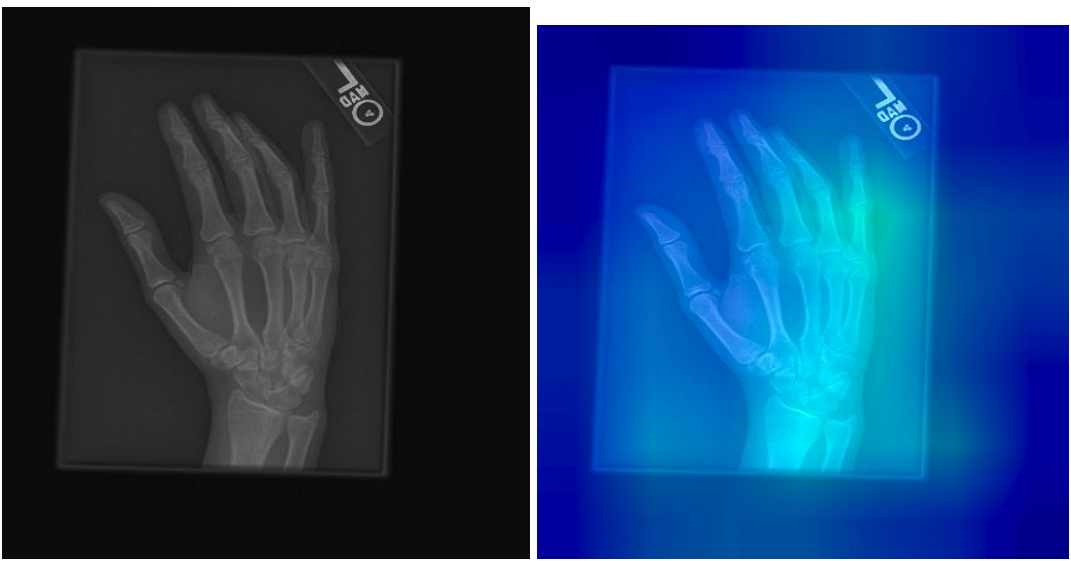

Figure 11. Normal Hand predicted using Ensemble200 as normal(negative, 1) with a score of (0.88). Image at the top is the input to the model, and the image at the bottom is the output of the Grad CAM algorithm with highlighted regions.

This figure below shows the results of running the predictions with a web based application and comparing the performance of three different models(Ensemble200, Single 1-MobileNet, Single2-NASNetMobile respectively)

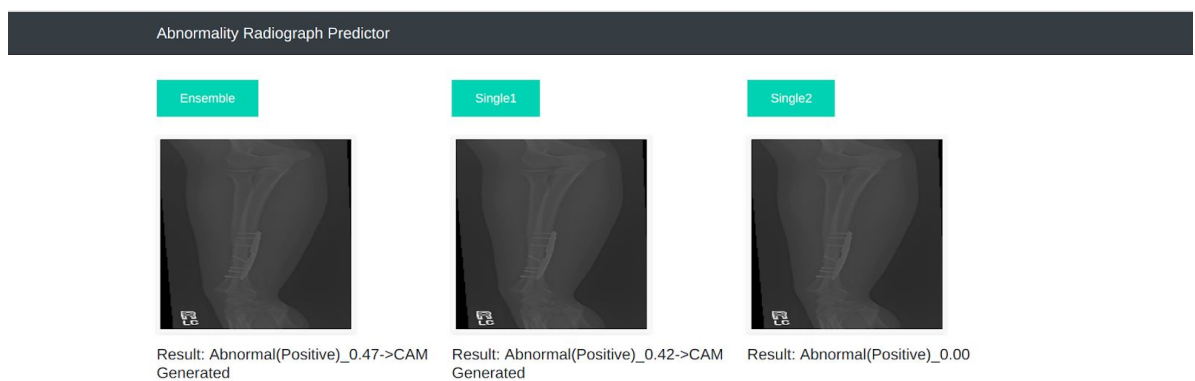

Figure 12. Abnormality Predictor Web App. Shows the prediction of an abnormal forearm as abnormal by all the three models and the Grad Class Activation Maps generated.

## 6. Conclusions

In this study, we demonstrated the use of ensembles in improving the overall performance and generalization of Convolutional Neural Networks (CNN) with other optimization techniques in predicting abnormality or normality in Musculoskeletal radiographs.

The evaluation results show most of the ensembles performance outperformed the single models, but were also instances where the single models outperformed some ensembles. This highlighted the combination of the varying architectures is also an important factor while building ensemble models together with other factors as the network depth, size, the total number of parameters. In the evaluation results, the best model, which was the ensemble200 [DenseNet201, MobileNet, NASNETmobile] which was built of one deep CNN network architecture of 201 layer deep, and two small size CNN architectures. The Kappa score of the best ensemble [ensemble200] was 0.66 compared with the best single model (MobileNet) with a kappa score of 0.41. Hence, this shows the use of ensembles with the MURA dataset has improved performance and generalization of the model.

In addition, other optimization techniques such class weight balancing was important to improve generalization of the classes as the dataset had more normal studies than abnormal studies. Use of data augmentation techniques as random lateral inversions and rotations and normalization of the images were important to avoid overfitting.

Use of an interactive web application in a clinical setup demonstrated the ease of integration of the abnormality predictor in the clinical workflow to support clinicians in their day to day. Application of the Grad Class Activation Maps in the clinical workflow would help in interpretation and validation of the results of the models as seen from the Grad CAM generated images with highlighted regions.

## Acknowledgement

We are very thankful to the Stanford Machine Learning Group in partnership with Stanford Hospital for making this project possible by publishing the Largest Musculoskeletal Radiographs dataset for further research of application of deep learning techniques in medical image diagnostics. The work by the research team led by Prof. Andrew Ng is admirable and should encourage more data sharing for modeling and ultimately for the common good.